# Molecular Decomposition of the Area Compressibility Modulus of Lipid Membranes

**Konstantin V. Pinigin**

Frumkin Institute of Physical Chemistry and Electrochemistry Russian Academy of Sciences, 31-4, Leninsky prospect, 119071, Moscow, Russia

Correspondence: piniginkv@gmail.com

**Abstract:** The area compressibility modulus is a key descriptor of lipid membrane mechanics, but the molecular interactions that determine this elasticity are not evident from the total modulus alone. Here, molecular dynamics simulations of a POPC bilayer described by the coarse-grained Martini 3 force field were combined with virial stress analysis to decompose the area compressibility modulus into contributions from molecular groups and their interactions. Partial lateral tensions were evaluated as functions of membrane area strain, and their derivatives at the equilibrium state were used to obtain the corresponding contributions to the area compressibility modulus. Hydrocarbon chains provided the dominant positive contribution, accounting for approximately 70% of the total modulus, whereas water–lipid interactions contributed approximately 30%, predominantly through water–headgroup interactions. Within the bilayer, the headgroup and headgroup–chain contributions were of similar magnitude but opposite in sign and nearly canceled. Direct intermonolayer interactions contributed only about 3% of the bilayer-only modulus, indicating an almost additive elastic response of the two monolayers. The hierarchy of elastic contributions differed markedly from that of equilibrium partial tensions, showing that interactions responsible for static stress balance are not necessarily those governing membrane stiffness. This decomposition provides a microscopic interpretation of lipid bilayer area elasticity.

## 1. Introduction

Biological membranes define cellular compartments, regulate molecular transport, and provide a dynamic platform for a wide range of biochemical processes [1,2]. Their function depends not only on composition and molecular organization but also on the ability of the lipid bilayer to sustain and redistribute mechanical stresses during deformation. Two central descriptors of this mechanical behavior are the lateral tension, $\gamma$, and the area compressibility modulus, $k_A$ [3,4]. Whereas lateral tension characterizes the mechanical state of the membrane at a given area, the area compressibility modulus measures the differential resistance of the bilayer to in-plane expansion or compression.

Area elasticity is important well beyond uniform stretching. Membrane bending necessarily produces opposite area strains across the bilayer thickness and is therefore intrinsically coupled to lateral compression and dilation of the monolayers [5–10]. The energetic cost associated with these deformations contributes to membrane-mediated interactions [11–14], pore formation [15–17], the mechanics of lipid-domain boundaries [18–26], and membrane fission and fusion [27–31]. At the continuum level, these effects are commonly described by a small set of elastic moduli. At the molecular level, however, the origin of the corresponding stiffness is much less transparent: it is not immediately evident which parts of a lipid molecule, or which lipid–solvent interactions, bear the mechanical response to a change in membrane area.

Molecular dynamics simulations provide a direct route to this microscopic description by resolving the stress generated by individual molecular interactions [32–38]. A key result of earlier stress analyses is that a tensionless bilayer need not be locally stress-free. Instead, mechanical equilibrium can arise from the cancellation of large positive and negative contributions associated with different regions and molecular components of the membrane [33,34]. Decompositions of the lateral tension into spatial, energetic, entropic, and molecular-group contributions have therefore been valuable for identifying how water, lipid headgroups, and hydrocarbon chains participate in the equilibrium stress balance [33,34]. These decompositions, however, characterize the static mechanical state and do not by themselves determine which molecular subsystems dominate the elastic response.

This distinction follows directly from the definition of the area compressibility modulus as the derivative of lateral tension with respect to relative area strain. A molecular subsystem may carry a large partial tension at equilibrium while its tension changes only weakly under deformation; conversely, a small equilibrium contribution may have a steep strain dependence and therefore make a large contribution to the modulus. Thus, the molecular origin of membrane stiffness is encoded not in the magnitude of each partial tension, but in its response to strain. Decomposing the strain-dependent tension and differentiating the resulting partial contributions provides a natural way to assign partial area compressibility moduli to specific structural groups and interactions, thereby separating the roles of chain packing, hydration, and intermonolayer coupling.

In the present study, this approach is applied to a POPC bilayer described with the coarse-grained Martini 3 force field. Virial stress analysis is used to partition the lateral tension into contributions from water, lipid headgroups, hydrocarbon chains, water-lipid interactions, and interactions between the two monolayers. The strain dependence of each contribution is then used to obtain the corresponding partial area compressibility modulus. The resulting decomposition shows that the hydrocarbon-chain group provides the dominant contribution to area elasticity, water–lipid interactions provide a substantial secondary contribution, and direct intermonolayer coupling contributes only weakly to the area modulus despite carrying a large equilibrium partial tension. By resolving equilibrium stress and differential stiffness within a common framework, the analysis provides a molecular picture of how membrane architecture and hydration determine the resistance of a lipid bilayer to changes in area.

## 2. Theory

### *2.1. Lateral tension*

Consider a planar lipid bilayer oriented in the *xy*-plane of a rectangular periodic simulation cell with instantaneous dimensions $L_x$, $L_y$, and $L_z$. The membrane area is $A = L_xL_y$ and the instantaneous simulation-cell volume is $V = AL_z$.

In molecular dynamics, the instantaneous pressure tensor can be written as the sum of kinetic and configurational contributions [39,40]

$$\mathbf{P} = \mathbf{P}^{kin} + \mathbf{P}^{conf}, \tag{1}$$

where the kinetic contribution is

$$\mathbf{P}^{kin} = \frac{1}{V}\sum_{\alpha}\frac{\mathbf{p}_\alpha \otimes \mathbf{p}_\alpha}{m_\alpha}, \tag{2}$$

and the configurational contribution is determined by the molecular virial tensor **W**,

$$\mathbf{P}^{conf} = \mathbf{W} / V. \tag{3}$$

The configurational virial has the classical force–position form

$$\mathbf{W} = \sum_{\alpha}\mathbf{r}_\alpha \otimes \mathbf{F}_\alpha, \tag{4}$$

where $\mathbf{r}_\alpha$ is the position of particle $\alpha$ and $\mathbf{F}_\alpha$ is the total configurational force acting on it.

For a planar membrane, the instantaneous lateral and normal pressures are

$$P_L = \left(P_{xx} + P_{yy}\right)/2, \quad P_N = P_{zz}. \tag{5}$$

The lateral tension is obtained from the anisotropy of the pressure tensor as [41]

$$\gamma = \left\langle L_z\left[P_N - P_L\right]\right\rangle \tag{6}$$

where angular brackets denote an ensemble or time average. The equilibrium membrane area $A_0$ is defined by $\gamma = 0$; expansion corresponds to $\gamma > 0$ and compression to $\gamma < 0$.

### *2.2. Area compressibility modulus*

The area compressibility modulus $k_A$ quantifies the differential resistance of the membrane to changes in area. It is defined as the derivative of the lateral tension with respect to the relative area strain [3,6,42],

$$k_A = \left.\frac{d\gamma}{d\varepsilon}\right|_{\varepsilon=0}, \tag{7}$$

where

$$\varepsilon = \frac{\left(A - A_0\right)}{A_0} \tag{8}$$

is the relative area strain. Thus, $k_A$ is the local slope of the lateral tension–area relation at the equilibrium state.

*2.3. Molecular decomposition of the virial pressure*

The configurational virial is linear in the interparticle forces and is therefore additive over the interaction terms contributing to the potential energy. Consider a potential energy represented as

$$U = \sum_{a} U_{a} , \tag{9}$$

where $U_a$ denotes an individual interaction contribution, which may involve two, three, or, more generally, an arbitrary number of particles. The contribution of $U_a$ to the force acting on particle α is

$$\mathbf{F}_{\alpha}^{(a)} = -\frac{\partial U_{a}}{\partial \mathbf{r}_{\alpha}} , \tag{10}$$

and its contribution to the configurational virial is

$$\mathbf{W}_{a} = \sum_{\alpha \in a} \mathbf{r}_{\alpha} \otimes \mathbf{F}_{\alpha}^{(a)} . \tag{11}$$

The total configurational virial is therefore

$$\mathbf{W} = \sum_{a} \mathbf{W}_{a} . \tag{12}$$

This representation does not require a many-body interaction to be rewritten as a set of effective pair forces. The total virial contribution of a many-body potential is well defined, whereas a further decomposition of that contribution into effective pair-force terms is generally not unique [39,43–45].

Suppose that the system is partitioned into disjoint molecular groups. Each interaction term is classified according to the set of groups represented among its participating particles. If all participating particles belong to a single group $i$, the term is assigned to the corresponding intragroup contribution $\mathbf{P}_{ii}$. If the participating particles belong to exactly two groups $i$ and $j$, the complete virial of the term is assigned to the corresponding intergroup contribution $\mathbf{P}_{ij}$. More generally, if an $n$-body interaction spans more than two groups, its complete virial is retained as a single multigroup contribution associated with all participating groups, rather than being distributed among pairwise group contributions. For the groupings used in the present study, every force-field interaction term involves particles from no more than two of the defined groups, so the notation $\mathbf{P}_{ii}$ and $\mathbf{P}_{ij}$ is sufficient.

For two disjoint groups $i$ and $j$ that together constitute the system under consideration, the pressure tensor can therefore be written as

$$\mathbf{P}_{i \cup j} = \mathbf{P}_{i} + \mathbf{P}_{j} + \mathbf{P}_{ij} . \tag{13}$$

The kinetic contribution is contained in the single-group terms, whereas the intergroup term is configurational. Because lateral tension is linear in the pressure-tensor components, the same decomposition applies to the tension. For the groupings considered here, in which all contributions are represented by $ii$ or $ij$ terms,

$$\gamma = \sum_{i} \sum_{j \geq i} \gamma_{ij} , \tag{14}$$

where the partial contribution associated with groups $i$ and $j$ is

$$\gamma_{ij} = \left\langle L_z \left[ P_{ij,zz} - \left( P_{ij,xx} + P_{ij,yy} \right) / 2 \right] \right\rangle \tag{15}$$

Terms with $i = j$ denote intragroup contributions, whereas terms with $i < j$ denote interactions between distinct groups. This construction can be applied successively at different levels of molecular resolution. Here, it is used to resolve contributions from water and the lipid bilayer, the two monolayers and their mutual interaction, lipid headgroups and hydrocarbon chains and their interaction, as well as water–headgroup and water–chain interactions.

*2.4. Decomposition of the area compressibility modulus*

Since the lateral tension is additive, its derivative with respect to area strain is additive as well. The area compressibility modulus can therefore be decomposed as

$$k_A = \sum_i \sum_{j \geq i} k_{A,ij} , \tag{16}$$

where each partial contribution is defined by

$$k_{A,ij} = \left. \frac{d\gamma_{ij}}{d\varepsilon} \right|_{\varepsilon=0} . \tag{17}$$

Thus, the contribution of a molecular group or interaction to the area compressibility modulus is determined by how its virial contribution to the lateral tension changes under area deformation. A large equilibrium partial tension does not necessarily imply a large elastic contribution, and conversely, a small equilibrium partial tension may make a large contribution to $k_A$ if it varies strongly with strain.

## 3. Materials and Methods

*3.1. System setup*

Molecular dynamics simulations were performed for a lipid bilayer composed of POPC (1-palmitoyl-2-oleoyl-sn-glycero-3-phosphocholine). POPC was selected as the model lipid because of its biological relevance and well-characterized structural properties. It is one of the most abundant and extensively studied phospholipids in biological membranes [46–48]. With one saturated palmitoyl (16:0) chain and one unsaturated oleoyl (18:1) chain, POPC has an intermediate degree of unsaturation and provides a representative model of a fluid-phase biological membrane, including realistic membrane thickness and lateral fluidity. It therefore serves as a well-established benchmark for studies of membrane mechanics.

Simulations were carried out using the coarse-grained Martini 3 force field [49]. The coarse-grained representation was chosen because the aim of the present study is to resolve membrane elasticity at the level of major molecular components and their interactions, rather than at the level of individual atomic degrees of freedom. Martini provides a particularly suitable representation for this purpose: chemically distinct molecular fragments are mapped onto a limited set of interaction sites whose nonbonded interactions are parametrized primarily against thermodynamic partitioning and transfer properties of representative chemical compounds [49,50]. This minimal molecular representation retains the distinction between hydrophobic, hydrophilic, and interfacial components that is central to the present decomposition, while avoiding atomistic detail that is not required for assigning the mechanical response to these molecular groups. The use of a coarse-grained model also enables the long simulations required for accurate derivatives of partial tensions with respect to membrane area.

A bilayer containing 256 POPC lipids (128 lipids per monolayer) was constructed using CHARMM-GUI [51–55]. The system was solvated with 11 water beads per POPC molecule, giving 2,816 water beads in total. Because one Martini 3 water bead represents four real water molecules, this corresponds to 11,264 water molecules.

### *3.2. Simulation parameters*

Molecular dynamics simulations were performed with GROMACS [56,57]. The Verlet cutoff scheme [58] was used, with the neighbor list updated every 20 steps and a Verlet-buffer tolerance of 0.005 kJ $mol^{-1}$ $ps^{-1}$. The equations of motion were integrated using the standard molecular dynamics integrator [59] with a time step of 0.02 ps. Nonbonded interactions were truncated at 1.1 nm. Electrostatic interactions were treated with the reaction-field method using a relative dielectric constant of 15, as prescribed by the Martini force field. Van der Waals interactions were treated with a plain cutoff and the potential-shift-Verlet modifier.

Temperature was maintained at 300 K using the velocity-rescale thermostat [60] with a coupling time of 1.0 ps. The membrane and solvent were coupled as separate temperature groups. Pressure was controlled with the Berendsen barostat [61], using a coupling time of 3.0 ps and a compressibility of $3 \times 10^{-5}$ $bar^{-1}$ in both the lateral and normal directions. The normal reference pressure was set to 1 bar, and surface-tension pressure coupling was used to impose the specified lateral tensions.

Simulations were performed at applied lateral tensions of −100, −50, 0, 75, and 150 bar nm. Each system was equilibrated for 100 ns, followed by a 3 μs production run. Trajectory frames were saved every 5 ps for subsequent calculation of the virial stress tensor.

### *3.3. Calculation of virial stress*

The virial stress tensor was calculated using GROMACS-LS [36,44,62,63]. In the sign convention adopted here, the stress tensor, **S**, is defined as the negative of the pressure tensor, **P**, introduced in Section 2, i.e., $\mathbf{S} = -\mathbf{P}$. GROMACS-LS implements the classical Irving–Kirkwood–Noll definition of microscopic stress [64,65] and performs spatial averaging on a three-dimensional rectangular grid using a trilinear weighting function. Because a spatially resolved stress profile was not required in the present study, GROMACS-LS was configured with a single grid cell encompassing the entire simulation box. Under this configuration, the calculated quantity is the virial stress tensor.

The virial stress was decomposed at several levels of structural resolution. At the coarsest level, the system was separated into water and the lipid bilayer. The bilayer contribution was further resolved into the upper and lower monolayers and their mutual interaction. The lipid contribution was also resolved into headgroup and hydrocarbon-chain components and their interaction. Finally, the water–lipid interaction was resolved into water–headgroup and water–chain contributions.

In Martini 3, each POPC molecule is represented by 12 beads: one choline bead, one phosphate bead, two beads representing the glycerol moiety, four beads representing the palmitoyl tail, and four beads representing the oleoyl tail. For the present decomposition, the lipid headgroup was defined as the four beads comprising choline, phosphate, and glycerol, whereas the remaining eight beads forming the two acyl tails were assigned to the hydrocarbon-chain group. Throughout the manuscript, the term "headgroup" includes the glycerol moiety, whereas "hydrocarbon chains" refers specifically to the hydrocarbon acyl tails.

To evaluate the stress contribution from a specific bead group, all beads not belonging to that group were removed from the saved molecular dynamics trajectory, and the stress tensor was calculated from the resulting reduced trajectory using the same procedure as for the complete system.

Accordingly, a single-group contribution contains the kinetic contribution of the beads belonging to that group together with configurational contributions from interaction terms whose participating beads all belong to that group. Any interaction involving at least one bead from another group is excluded, including both nonbonded interactions and bonded terms such as bonds and angle potentials. For two disjoint groups, the intergroup stress contribution was obtained by subtracting the stress tensors of the two individual groups from that of the combined system containing both groups. The kinetic contributions cancel in this subtraction, so the resulting intergroup contribution is purely configurational.

### *3.4. Calculation of partial tensions and partial compressibility moduli*

Partial lateral tensions were obtained from the corresponding partial stress tensors according to Eq. (6), taking into account the stress convention $\mathbf{S} = -\mathbf{P}$ defined in Section 3.3. For statistical analysis, the trajectory at each applied lateral tension was divided into 20 equal-length blocks. These block values were then resampled with replacement (bootstrapping) to estimate statistical uncertainties.

Partial area compressibility moduli were determined from the strain dependence of the corresponding partial tensions. For each contribution, the partial tension–strain relation was fitted by least squares to a quadratic polynomial, and the coefficient of the linear term was taken as the derivative at zero strain. To obtain robust mean values and uncertainties, the fitting procedure was repeated 1,000 times. In each bootstrap iteration, one partial-tension estimate was generated at each strain value by resampling the 20 block values, after which the quadratic fit was applied to the resulting set of points. The distribution of fitted linear coefficients was then used to determine the reported partial modulus and its statistical uncertainty.

## 4. Results

### *4.1. Equilibrium partial tensions*

Table 1 reports the equilibrium partial lateral tensions, $\gamma_{ij}$, together with the corresponding partial area compressibility moduli, $k_{A,ij}$. Although the total lateral tension is zero by construction at the reference state ($-0.004 \pm 0.011$ $k_BT$ nm$^{-2}$), the individual molecular contributions are not small. Instead, mechanical equilibrium results from cancellation among sizeable positive and negative partial tensions, and this cancellation can be verified independently for several partitions of the system.

Within the lipid subsystem, the upper and lower monolayers carry positive partial tensions of $3.203 \pm 0.011$ and $3.200 \pm 0.010$ $k_BT$ nm$^{-2}$, respectively. The remaining contribution required to recover the bilayer-only tension is assigned to the monolayer–monolayer interaction and amounts to $-9.046 \pm 0.003$ $k_BT$ nm$^{-2}$. Thus, although each monolayer separately carries a positive partial tension, the intermonolayer interaction contribution makes the total bilayer-only tension negative.

The same bilayer-only contribution can alternatively be expressed in terms of the headgroup, hydrocarbon-chain, and headgroup-chain terms. Their equilibrium contributions are $11.585 \pm 0.008$, $-3.280 \pm 0.010$, and $-10.948 \pm 0.005$ $k_BT$ nm$^{-2}$, respectively, and sum to the bilayer-only value within rounding, as required by the additive construction of the decomposition. Thus, the leaflet-based and molecular-group partitions provide two alternative representations of the same bilayer contribution.

A second cancellation occurs between the lipid and solvent sectors. Water alone contributes $6.841 \pm 0.010$ $k_BT$ nm$^{-2}$, while water-headgroup and water-chain interactions contribute $-2.422 \pm 0.009$ and $-1.779 \pm 0.002$ $k_BT$ nm$^{-2}$, respectively. The combined solvent-related equilibrium contribution is therefore approximately $+2.640$ $k_BT$ nm$^{-2}$, which compensates the bilayer-only value of approximately $-2.643\ k_BT\,\mathrm{nm}^{-2}$. Consequently, the near-zero total tension of the complete system results from compensation among large internal stresses rather than from each molecular subsystem being individually tensionless.

The magnitudes of the equilibrium partial tensions are distributed very unevenly among molecular terms. The largest single-group contribution is the positive headgroup tension, while the headgroup-chain and monolayer-monolayer interactions carry comparably large negative tensions. Water also makes a substantial positive contribution. This ordering differs strongly from the hierarchy of the corresponding elastic derivatives discussed below.

**Table 1.** Partial lateral tensions, $\gamma_{ij}$, evaluated at zero total lateral tension, and the corresponding partial area compressibility moduli, $k_{A,ij}$. In the present grouping, the headgroup contains the choline, phosphate, and glycerol beads, whereas the chain group contains only the hydrocarbon acyl-tail beads. Uncertainties are standard deviations.

| Contribution | Partial lateral tension, $\gamma_{ij}$ ($k_BT$ nm$^{-2}$) | Partial area compressibility modulus, $k_{A,ij}$ ($k_BT$ nm$^{-2}$) |
|---|---|---|
| Total | −0.004 ± 0.011 | 71.46 ± 0.27 |
| Bilayer-only | −2.643 ± 0.012 | 50.05 ± 0.23 |
| Water | 6.841 ± 0.010 | 0.70 ± 0.19 |
| Headgroup | 11.585 ± 0.008 | −15.02 ± 0.11 |
| Hydrocarbon chains | −3.280 ± 0.010 | 50.67 ± 0.21 |
| Water-chain | −1.779 ± 0.002 | −1.37 ± 0.05 |
| Water-headgroup | −2.422 ± 0.009 | 22.07 ± 0.17 |
| Headgroup-chain | −10.948 ± 0.005 | 14.41 ± 0.15 |
| Upper monolayer | 3.203 ± 0.011 | 24.45 ± 0.18 |
| Lower monolayer | 3.200 ± 0.010 | 24.10 ± 0.22 |
| Monolayer-monolayer | −9.046 ± 0.003 | 1.51 ± 0.07 |

### *4.2. Extraction of partial area compressibility moduli*

The partial area compressibility moduli were obtained from the strain dependence of the corresponding partial tensions according to Eq. (17). For each contribution, $\gamma_{ij}(\varepsilon)$ was fitted with a quadratic polynomial as described in Section 3.4. The quantity $k_{A,ij}$ is the derivative at $\varepsilon = 0$ and therefore equals the coefficient of the linear term of the fit. Retaining the quadratic term allows the simulated strain dependence to be represented over the full sampled range while defining the reported modulus locally at the undeformed reference state.

Figure 1 shows the hydrocarbon-chain contribution as a representative example. Its partial tension changes smoothly across the simulated strain range, and the derivative of the quadratic fit at the origin gives $k_{A,ij} = 50.67 \pm 0.21\ k_BT\,\mathrm{nm}^{-2}$. The other partial tensions were analyzed in the same manner, so that all entries in the third column of Table 1 are obtained from the same derivative-based definition rather than from the magnitudes of the corresponding equilibrium tensions.

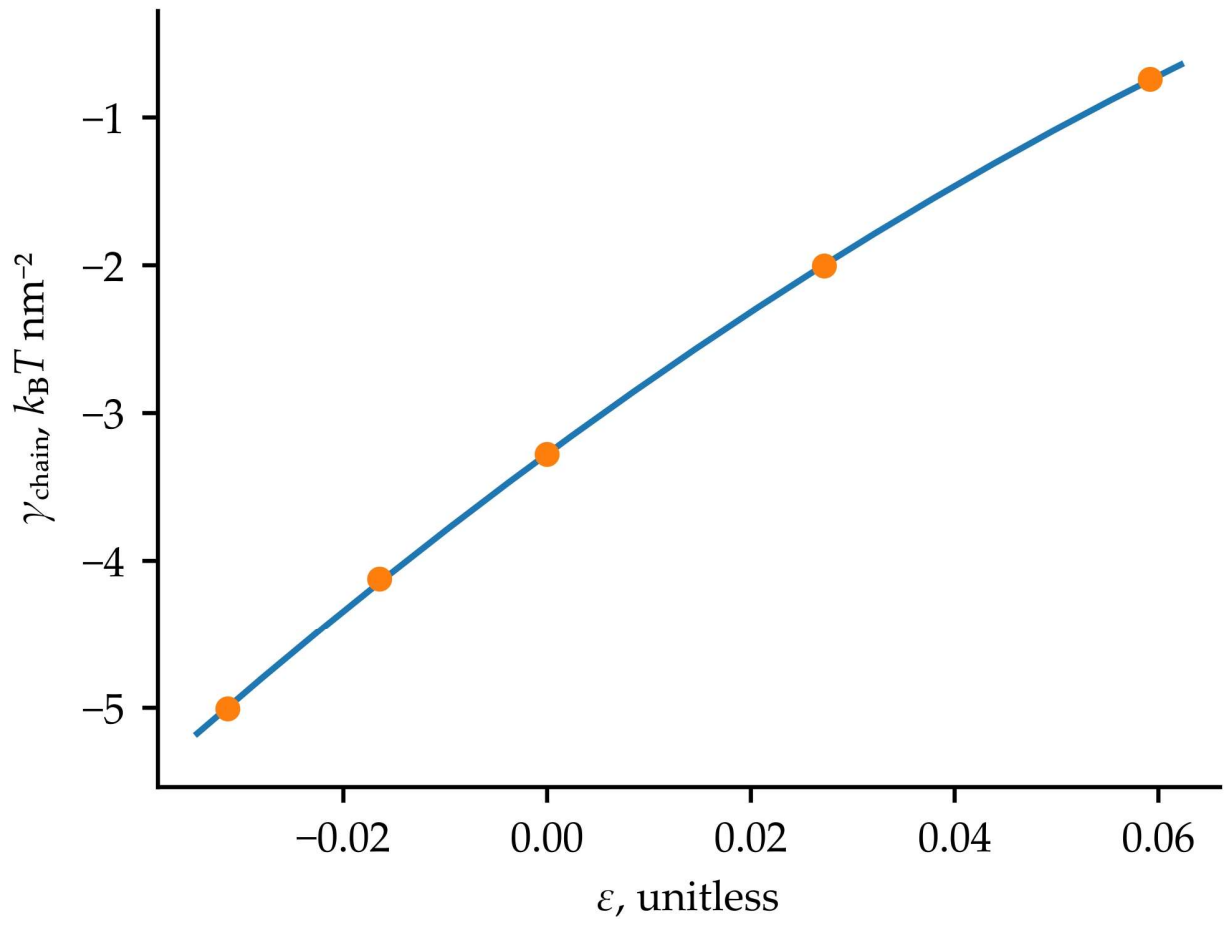

**Figure 1.** Partial lateral tension associated with the chain contribution as a function of relative area strain $\varepsilon$. The solid curve is the quadratic fit used to determine $k_{A,\text{chain}}$ from the derivative at $\varepsilon = 0$. Statistical uncertainties are smaller than the plotting symbols.

### *4.3. Molecular-group hierarchy of area elasticity*

The molecular-group decomposition displays a pronounced hierarchy of elastic contributions (Table 1 and Figure 2). The hydrocarbon-chain group provides the dominant positive term, 50.67 ± 0.21 $k_BT$ nm$^{-2}$, corresponding to 70.9% of the total modulus, 71.46 ± 0.27 $k_BT$ nm$^{-2}$. This contribution alone is essentially equal to the complete bilayer-only modulus of 50.05 ± 0.23 $k_BT$ nm$^{-2}$. The remaining bilayer-internal terms are large but strongly compensating: the headgroup contribution is −15.02 ± 0.11 $k_BT$ nm$^{-2}$, whereas the headgroup-chain interaction contributes +14.41 ± 0.15 $k_BT$ nm$^{-2}$. Their sum is close to zero, and adding both terms to the hydrocarbon-chain contribution gives 50.06 $k_BT$ nm$^{-2}$, reproducing the bilayer-only result within rounding.

Water-lipid interactions form the second major positive contribution, but they are highly asymmetric between the two lipid groups. The water-headgroup term is 22.07 ± 0.17 $k_BT$ nm$^{-2}$, whereas the water-chain contribution is slightly negative, −1.37 ± 0.05 $k_BT$ nm$^{-2}$. Their directly evaluated combined contribution is 20.70 ± 0.15 $k_BT$ nm$^{-2}$, or 29.0% of the total area compressibility modulus. Thus, in the present grouping, the elastic effect of membrane hydration is concentrated predominantly in interactions between water and the headgroup region, which includes the glycerol moiety.

The direct water contribution is small, 0.70 ± 0.19 $k_BT$ nm$^{-2}$. Adding this term to the water-lipid interaction contribution gives approximately 21.40 $k_BT$ nm$^{-2}$, or 30.0% of the total modulus. Together with the bilayer-only contribution, these terms account for the total modulus within rounding, as expected from the additive decomposition.

The two largest positive elastic contributions are consequently associated with the hydrocarbon chains and water-headgroup interactions. Several other terms act mainly as compensating corrections: the negative headgroup contribution is largely offset by the positive headgroup-chain term, while the small negative water-chain term partially reduces the water-headgroup contribution. Figure 2 summarizes this hierarchy in cumulative form.

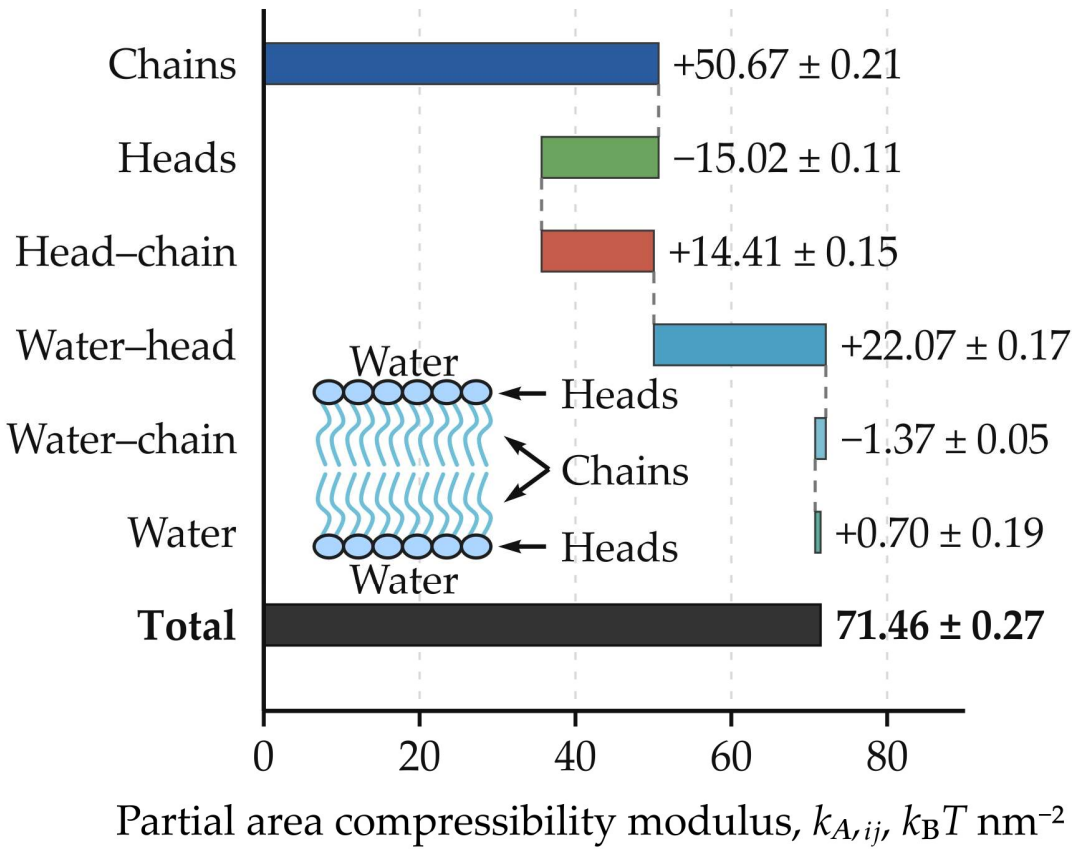

**Figure 2.** Molecular decomposition of the area compressibility modulus, $k_A$, of the POPC membrane system. The cumulative bars show the hydrocarbon-chain, headgroup, headgroup–chain, water–headgroup, water–chain, and water contributions. Numerical labels give the corresponding values and statistical uncertainties in $k_B T$ nm$^{-2}$.

### *4.4. Leaflet decomposition and intermonolayer coupling*

The leaflet decomposition provides a complementary view of the bilayer-only elastic response (Figure 3). The upper and lower monolayers contribute 24.45 ± 0.18 and 24.10 ± 0.22 $k_B T$ nm$^{-2}$, respectively, differing by only 0.35 $k_B T$ nm$^{-2}$. Their sum, 48.55 $k_B T$ nm$^{-2}$, already accounts for approximately 97% of the bilayer-only modulus. The remaining 1.51 ± 0.07 $k_B T$ nm$^{-2}$, or about 3%, is assigned to the monolayer–monolayer interaction. Adding all three contributions gives 50.06 $k_B T$ nm$^{-2}$, in agreement with the bilayer-only value of 50.05 ± 0.23 $k_B T$ nm$^{-2}$ within rounding. Thus, the two monolayers respond almost additively to uniform area deformation, with direct intermonolayer coupling providing only a small correction.

This elastic hierarchy contrasts sharply with the equilibrium-tension decomposition. The same monolayer–monolayer interaction carries a large negative equilibrium tension, −9.046 ± 0.003 $k_B T$ nm$^{-2}$, and is essential for converting the two positive monolayer tensions into the negative bilayer-only tension. Thus, direct intermonolayer interactions are important for the static stress balance but weak for the differential area stiffness.

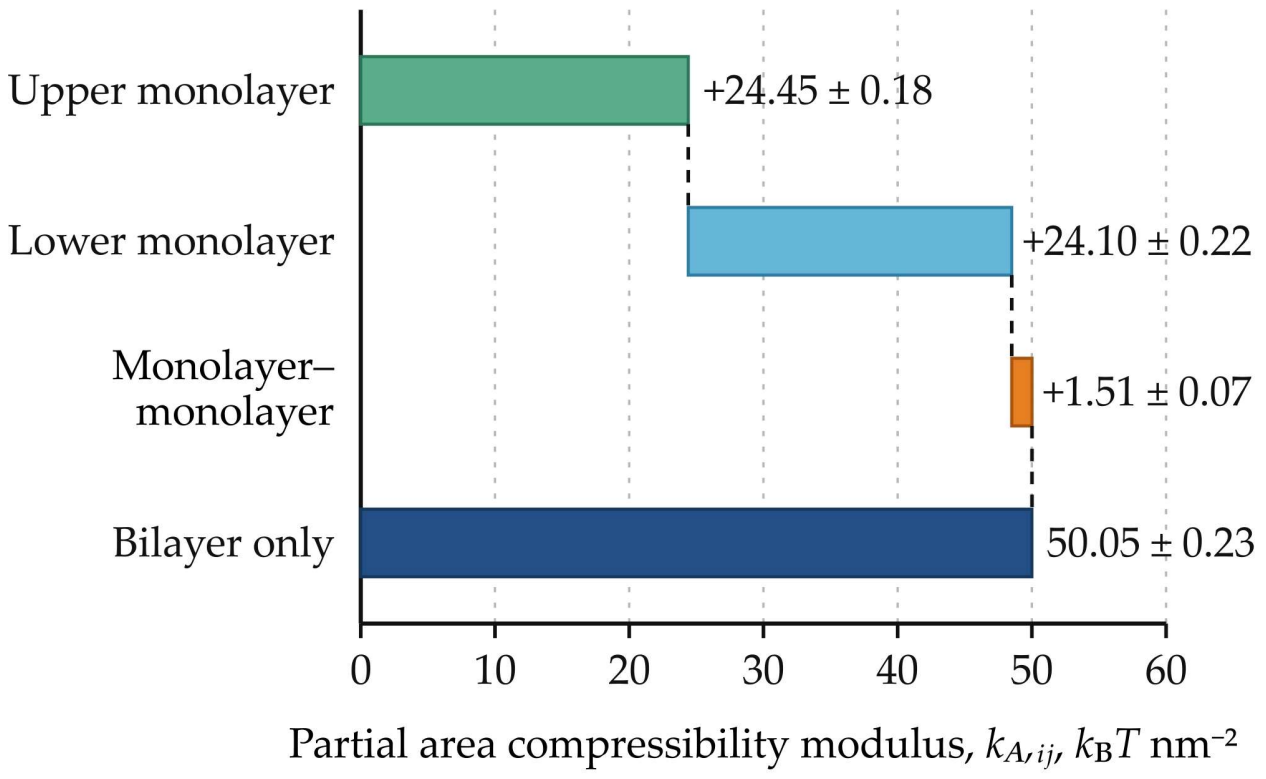

**Figure 3.** Leaflet decomposition of the bilayer-only area compressibility modulus. The upper- and lower-monolayer terms nearly exhaust the bilayer-only modulus, while the direct monolayer-monolayer interaction provides only a small correction. Dashed connectors indicate the cumulative sum; numerical labels give values and statistical uncertainties in $k_B T$ nm$^{-2}$.

*4.5. Equilibrium stress versus differential stiffness*

The equilibrium partial tensions and elastic contributions show markedly different hierarchies (Figure 4). The hydrocarbon-chain term has a moderate negative equilibrium tension of −3.280 ± 0.010 $k_BT$ nm$^{-2}$, yet it supplies the dominant elastic contribution of 50.67 ± 0.21 $k_BT$ nm$^{-2}$. Water-headgroup interactions display a similar contrast: their equilibrium tension is only −2.422 ± 0.009 $k_BT$ nm$^{-2}$, whereas their contribution to $k_A$ reaches 22.07 ± 0.17 $k_BT$ nm$^{-2}$.

Conversely, several of the largest equilibrium stresses contribute little or even negatively to the area modulus. The monolayer-monolayer interaction carries −9.046 ± 0.003 $k_BT$ nm$^{-2}$ at equilibrium but contributes only 1.51 ± 0.07 $k_BT$ nm$^{-2}$ to $k_A$. Water carries a large positive equilibrium tension of 6.841 ± 0.010 $k_BT$ nm$^{-2}$, while its direct elastic contribution is only 0.70 ± 0.19 $k_BT$ nm$^{-2}$. Most strikingly, the headgroup term has the largest positive equilibrium partial tension, 11.585 ± 0.008 $k_BT$ nm$^{-2}$, but a negative elastic contribution of −15.02 ± 0.11 $k_BT$ nm$^{-2}$.

The headgroup-chain interaction provides the complementary example: it carries a large negative equilibrium tension of −10.948 ± 0.005 $k_BT$ nm$^{-2}$, but its elastic contribution is positive, 14.41 ± 0.15 $k_BT$ nm$^{-2}$. Thus, both the magnitude and the sign of $\gamma_{ij}$ at $\varepsilon = 0$ can be unrelated to the magnitude and sign of $k_{A,ij}$. A negative partial modulus contribution simply means that the corresponding partial tension decreases with increasing area at the reference state; it does not imply that the complete membrane is mechanically unstable.

Figure 4 visualizes this lack of one-to-one correspondence. Terms that lie far from zero on the equilibrium-tension axis can remain close to zero on the elastic axis, as for water and the monolayer-monolayer interaction, whereas the hydrocarbon-chain and water-headgroup terms provide the two largest positive elastic responses despite much smaller equilibrium tensions. The relevant quantity for area elasticity is therefore the local slope $d\gamma_{ij}/d\varepsilon$ at the reference state, not the equilibrium value of $\gamma_{ij}$ itself.

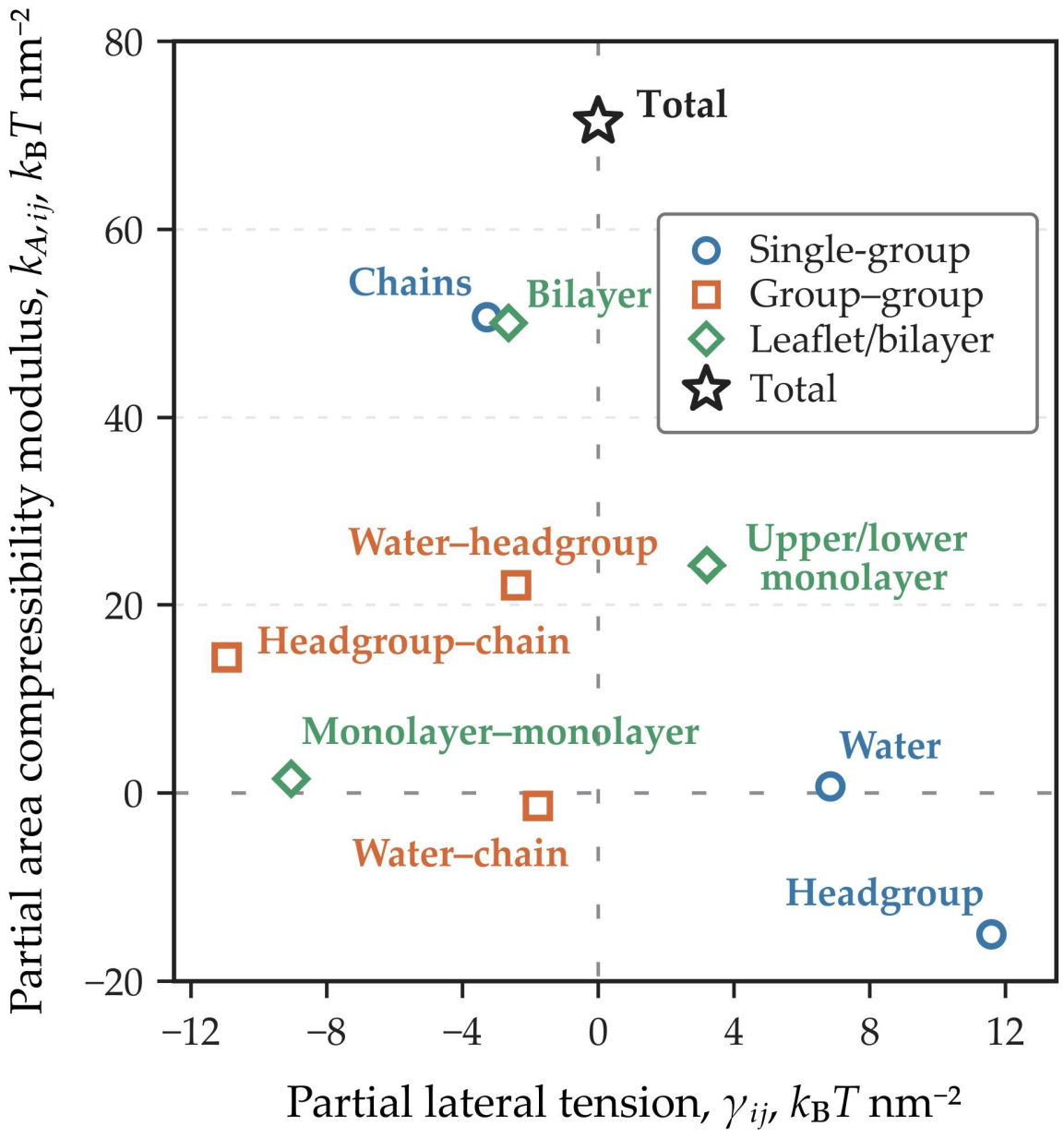

**Figure 4.** Relationship between the equilibrium partial lateral tension, $\gamma_{ij}$, and the corresponding partial contributions to the area compressibility modulus, $k_{A,ij}$. Each point corresponds to a term listed in Table 1; the upper and lower monolayers are represented by a single marker because their values nearly coincide. Marker shapes and colors distinguish single-group contributions, group-group interactions, leaflet/bilayer contributions, and the total system. Dashed lines indicate $\gamma_{ij} = 0$ and $k_{A,ij} = 0$. Statistical uncertainties are smaller than the plotting symbols.

Taken together, the decompositions establish four quantitative features of the POPC area-compressibility response: hydrocarbon chains dominate the positive elastic response; water-lipid interactions provide a substantial secondary contribution that is concentrated in water-headgroup interactions; the two monolayers respond almost additively with weak direct intermonolayer coupling; and the hierarchy of elastic contributions differs fundamentally from the hierarchy of equilibrium partial tensions. The molecular-group, solvent/bilayer, and leaflet partitions provide alternative additive representations of the same mechanical response.

## 5. Discussion

The present study extends previous analyses of equilibrium partial tensions by examining the strain dependence of each molecular contribution to lateral tension. Earlier stress decompositions showed that a tensionless bilayer can contain large positive and negative internal contributions that compensate one another [33,34]. However, the magnitude of a partial tension does not indicate how strongly that molecular subsystem resists a change in membrane area. By differentiating each partial tension with respect to area strain, the present approach separates these two mechanical roles: carrying static lateral stress and contributing to the area compressibility modulus. This comparison makes the distinction especially clear, because several molecular contributions change both rank and sign when equilibrium partial tensions are compared with their strain derivatives.

The total area compressibility modulus is 71.46 ± 0.27 $k_B T$ nm$^{-2}$, corresponding to approximately 296 mN m$^{-1}$ at 300 K. This value is somewhat higher than several earlier experimental estimates for fluid POPC membranes, which lie around 200–215 mN m$^{-1}$ [66–68], but is closer to the more recent values of 249 ± 35 and 255 ± 31 mN m$^{-1}$ reported by Drabik et al. [69]. The spread among these measurements shows that experimentally determined values of the POPC area compressibility modulus vary appreciably between studies. The present value is therefore somewhat above the central experimental estimates but remains comparable in magnitude to the measured stiffness of fluid POPC membranes. Within this overall elastic response, the hydrocarbon-chain contribution, 50.67 ± 0.21 $k_B T$ nm$^{-2}$, accounts for 70.9% of the total modulus and is therefore the dominant positive term. This result shows that the principal resistance to uniform lateral area deformation is associated with the acyl-tail region itself: the hydrocarbon chains alone provide a contribution essentially equal to the complete bilayer-only modulus.

This near equality arises because the remaining bilayer-internal contributions largely compensate one another. The headgroup-only contribution is large and negative, −15.02 ± 0.11 $k_B T$ nm$^{-2}$, whereas the headgroup-chain interaction contributes +14.41 ± 0.15 $k_B T$ nm$^{-2}$. These two terms nearly cancel, leaving the hydrocarbon-chain contribution as the dominant net source of bilayer stiffness. The negative headgroup contribution should be interpreted as a negative slope of the headgroup partial tension with area at the reference state, rather than as an instability of an isolated headgroup subsystem. Partial moduli are components of an additive decomposition of the total response and need not be positive individually.

Hydration provides the second major contribution. Water-headgroup interactions contribute 22.07 ± 0.17 $k_B T$ nm$^{-2}$, whereas water-chain interactions contribute only −1.37 ± 0.05 $k_B T$ nm$^{-2}$. Their combined water-lipid contribution is 20.70 ± 0.15 $k_B T$ nm$^{-2}$, or 29.0% of the total modulus. In contrast, the direct water term is only 0.70 ± 0.19 $k_B T$ nm$^{-2}$. Thus, the mechanical role of the solvent is expressed primarily through its interaction with the membrane rather than through the water subsystem alone, and in the present grouping this effect is localized predominantly to the headgroup-water interface. Because the headgroup contains the glycerol moiety as well as the choline and phosphate beads, this contribution represents the broader hydrated interfacial region rather than only the charged beads.

The leaflet decomposition reveals a similarly useful distinction between static stress and elastic response. The upper and lower monolayers contribute 24.45 ± 0.18 and 24.10 ± 0.22 $k_B T$ nm$^{-2}$, respectively. Their sum, 48.55 $k_B T$ nm$^{-2}$, already accounts for about 97% of the bilayer-only modulus, whereas the direct monolayer-monolayer interaction contributes only 1.51 ± 0.07 $k_B T$ nm$^{-2}$. This small elastic contribution is especially notable because the same interaction carries a large equilibrium partial tension of −9.046 ± 0.003 $k_B T$ nm$^{-2}$. The two monolayers are therefore nearly additive with respect to uniform area compression and stretching, even though their mutual interaction is crucial to the internal tension balance of the equilibrium bilayer.

More generally, Figure 4 demonstrates that equilibrium partial tension is not a proxy for elastic stiffness. The headgroup term provides the largest positive equilibrium tension but a negative contribution to $k_A$; the headgroup-chain interaction exhibits the opposite sign pattern; the hydrocarbon chains have a comparatively moderate equilibrium tension but dominate the modulus; and water carries a sizeable equilibrium stress while contributing very little directly to the stiffness. The relevant quantity for elasticity is therefore not the magnitude of a partial stress at the reference state, but how rapidly that stress changes when the membrane is strained. This distinction is important when interpreting molecular stress decompositions: the interactions that maintain mechanical equilibrium need not be the same interactions that dominate the energetic cost of deformation.

The decomposition also provides a natural framework for examining how membrane composition or environment redistributes elasticity. Perturbations that modify acyl-chain packing would be expected to act strongly on the dominant hydrocarbon-chain term, whereas changes in hydration, headgroup chemistry, or ionic conditions should be expressed primarily through the water-headgroup and headgroup-related contributions. Chain-ordering additives such as cholesterol are therefore a particularly natural test case for determining whether changes in the total modulus can be traced to a redistribution among these microscopic elastic channels. Such predictions, however, require explicit simulations of the modified systems rather than direct extrapolation from the present POPC bilayer.

Area compression and stretching are also intrinsically coupled to membrane bending because bending produces opposite area strains across the monolayer thickness [5–10]. The present results therefore suggest that the hydrocarbon-chain region and the hydrated headgroup interface are likely to be important contributors to bending elasticity as well. Nevertheless, partial contributions to $k_A$ cannot be transferred directly to a decomposition of the bending modulus: bending probes a spatially nonuniform strain field and depends on how the mechanical response is distributed across the membrane thickness. A quantitative decomposition of bending elasticity will consequently require an analysis specifically formulated for bending that explicitly accounts for the variation of strain across the membrane thickness.

The numerical values reported here are specific to the present POPC system, the Martini 3 model, and the chosen partition of the lipid beads. In particular, the present definition assigns the glycerol beads to the headgroup and reserves the term hydrocarbon chains for the acyl-tail beads only. The individual partial contributions necessarily depend on the chosen partition, while the additive relations ensure that the corresponding components reconstruct the same bilayer-only or total response.

## 6. Conclusions

By decomposing the strain-dependent lateral tension into molecular-group and interaction contributions, this study provides a microscopic interpretation of the area compressibility of a POPC membrane. The hydrocarbon-chain region provides the dominant positive contribution, accounting for approximately 70% of the total area compressibility modulus, while water–lipid interactions contribute approximately 30%, primarily through water–headgroup interactions. Within the bilayer, the negative headgroup contribution is nearly compensated by the positive headgroup–chain term, leaving the hydrocarbon chains as the principal source of bilayer stiffness. The two monolayers respond almost additively, with direct intermonolayer interactions contributing only about 3% of the bilayer-only modulus.

Comparison with the equilibrium partial tensions further shows that the molecular interactions that maintain the static stress balance are not necessarily those that govern differential stiffness. Thus, the strain dependence of partial stresses contains mechanical information that is not available from an equilibrium stress decomposition alone. The present framework can be extended to other lipid compositions and membrane environments to determine how changes in molecular packing, hydration, and interfacial interactions redistribute membrane elasticity.

**Funding:** The work was supported by the Ministry of Science and Higher Education of the Russian Federation.